\documentstyle[12pt,epsf]{article}
\def\be{\begin{equation}}
\def\ee{\end{equation}}
\def\H{{\hat H}}
\def\ffrac#1#2{\textstyle{#1\over#2}\displaystyle}
\gdef\journal#1, #2, #3, 1#4#5#6{{\sl #1~}{\bf #2}, #3, 1#4#5#6}

\begin{document}
\pagestyle{empty}
\vspace{-1mm}
\begin{flushright}
NSF-ITP-97-049\\
cond-mat/9705137
\end{flushright}
\vspace{5mm}
\begin{center}
{\bf \LARGE The Number of Incipient Spanning Clusters\\
 in Two-Dimensional Percolation\\}
\vspace{8mm}
{\bf \large John Cardy\\}
\vspace{2mm}
{Department of Physics\\
Theoretical Physics\\1 Keble Road\\Oxford OX1 3NP, UK\\
\& All Souls College, Oxford\\}
\end{center}
\vspace{6mm}
\begin{abstract}
Using methods of conformal field theory, we conjecture an exact form
for the probability that $n$ distinct clusters span a large rectangle
or open cylinder of aspect ratio $k$, in the limit when $k$ is large. 
\end{abstract}
\newpage

\pagestyle{plain}
\setcounter{page}{1}
\setcounter{equation}{0}
The study of the
structure of large clusters at the percolation threshold continues
to pose interesting problems whose solution sheds light on the nature of
the critical state
in general. Recently some attention has been paid to \em incipient
spanning clusters \em\ (ISCs). These are clusters which connect two disjoint
segments of the boundary of a macroscopically large region. 
Langlands et al.~\cite{Langlands} conjectured that the probability that at
least one such cluster exists (that is, that the segments are
connected) is invariant under conformal transformations. This statement
was placed in the context of conformal field theory in
Ref.~\cite{Cardy}, where an explicit formula was given for this crossing
probability. (For a review of the status of conformal invariance in
critical percolation, see Ref.~\cite{A1}.)

More recently, Aizenman \cite{A2} has considered, among other things, the
probability that there exist $n$ distinct ISCs connecting the two
segments.\footnote{One may distinguish the probability of
exactly $n$ ISCs from that of at least $n$. However, in the
limits considered in this note, these will turn out to be asymptotically
the same.} In the case of a rectangular region, $[0,kL]\times[0,L]$, he
has proved that the probability $P(n,k,L)$ that the strip is
traversed (in the direction of length $kL$) by $n$ independent clusters
satisfies the bounds
\be\label{aa}
Ae^{-\alpha n^2k}\leq P(n,k,L)\leq e^{-\alpha'n^2k},
\ee
where $\alpha$ and $\alpha'$ are (different) constants.
Note that, on the basis of scale invariance at the critical
point, $P(n,k,L)$ is expected to have a finite limit as $L\to\infty$.

In this note we extend the arguments of Ref.~\cite{Cardy} to determine
the exact behaviour of the scaling limit of $P$ for large $k$, namely that
\be\label{bb}
\lim_{L\to\infty}\ln P(n,k,L)\sim -{2\pi\over3}\,n(n-\ffrac12)\,k,
\ee
as $k\to\infty$ for any $n$. 

An analogous problem may be posed on a open-ended
cylinder of circumference $L$
and length $kL$. In this case we find, for $n\geq2$,
\be\label{bbc}
\lim_{L\to\infty}\ln P(n,k,L)\sim -{2\pi\over3}\,(n^2-\ffrac14)\,k,
\ee

In passing, we note that, as observed by Aizenman \cite{A2}, the form of
the result in (\ref{bb}-\ref{bbc}) is not surprising, even though it contradicts
what appears to have been a former consensus that only one such cluster should
exist which has only recently been
challenged and corrected by numerical evidence \cite{Sen,HuLin}.
For a recent review see Stauffer \cite{Stauffer}.
Indeed, if one imagines dividing the rectangle into two equal rectangles
each of size $[0,kL]\times[0,L/2]$, and assumes that the dominant event
will be that of approximately $n/2$ clusters spanning each half, then,
up to prefactors,
\be\label{pp}
P(n,k,L)\sim P((n/2), 2k, L/2)^2.
\ee
Together with the expected exponential dependence on $kL$ at fixed $L$, this
leads to the above form. A similar argument then may be made for the 
cylindrical geometry.

Our argument for the exact coefficients in (\ref{bb}-\ref{bbc}) is based on the
well-known mapping of bond percolation to the $q\to1$ limit of the $q$-state
Potts model, and the understanding of the critical theory of this model
through conformal field theory. In the Potts model, spins $s_i$ are
placed at the sites $i$ of a lattice, each taking one of $q$ possible
states. The partition function has the form ${\rm Tr}\,\prod_{ij}\big(1-p
+p\delta_{s_is_j}\big)$, where the product is over all links of the
lattice. This may be expanded in powers of $p/(1-p)$
so that each term corresponds to a particular realisation of bond percolation,
weighted by a factor of $q$ for each connected cluster. The limit
$q\to1$ then weights these as in percolation, but, as will be seen, it is
also often helpful to consider first the case of more general $q$.

In the rectangular geometry described above, it is useful
to express things in terms of the transfer matrix $e^{-\H(L)}$ for a
strip of width $L$. The partition function for the Potts model with
particular boundary
conditions at either end of a strip of finite length $kL$ then has the
form
\be\label{dd}
\langle A|e^{-kL\H(L)}|A\rangle,
\ee
where $|A\rangle$ is a \em boundary state \em
corresponding to the boundary conditions chosen. The symmetry of the
Potts model ensures that the degenerate subspaces of eigenstates of
$\H(L)$ may be chosen to transform according to irreducible
representations of the permutation group $S_q$ of $q$ objects.
Conformal field theory also asserts that, in the scaling limit, 
the states in the 
low-lying spectrum of $\H(L)$ transform
according to highest weight representations of a Virasoro algebra, and
their corresponding eigenvalues have the form $\pi(x+{\rm integer})/L$,
where $x$ is the highest weight. Thus (\ref{dd}) may also be written
\be\label{ee}
\sum_Re^{-\pi x_Rk}\sum_N\langle A|N\rangle\langle N|A\rangle
e^{-\pi Nk},
\ee
where the first sum is over highest weight representations $R$, and the
second over the states $|N\rangle$ is each representation.
(This notation is a little corrupt because there are in general many
states at level $N$.)

The simplest non-trivial irreducible 
representation of $S_q$ has dimension
$q-1$, corresponding to a vector $(\varphi_1,\ldots,\varphi_q)$ with 
$\sum_{a=1}^q\varphi_a=0$. An example is the Potts order parameter
$\varphi_a=\delta_{s_i,a}-q^{-1}$. Out of this other representations
may be built by taking
direct products. For example symmetric tensors $\varphi_{ab}$ with
$a\not=b$ and $\sum_a\varphi_{ab}=0$ give a representation of dimension
$(q-1)(q-2)/2$. In general we may construct tensors $\varphi_{ab\ldots}$
with $n$ components, none of whose indices are equal. Let us denote by
$R_n$ the Virasoro representation which also carries this representation of 
$S_q$ and which has the smallest weight $x_R$. Denote this weight by
$x_n$.

Suppose now that we are interested in those configurations in which
at least $n$ distinct ISCs connect the two ends of the strip. In that case it
is possible to colour these clusters with $n$ different colours of the
Potts model, and therefore
the states which propagate along the strip must carry
at least $n$ different colours. 
If there are fewer than $n$ ISCs, it is not possible to
make such an assignment. In the limit of large $k$, then, the partition
sum in (\ref{dd}) will be dominated by those state(s) transforming
according to representations $R_{n'}$ with $n'\geq n$. As we shall
argue, the highest weights $x_{n'}$ are monotonically increasing in $n'$.
Thus the states with $n'=n$ dominate the sum. 

What is the value of $x_n$? For $n=1$ the answer is known, since it
corresponds to the scaling dimension of the Potts order parameter near the
boundary of a semi-infinite system. It was conjectured in
Ref.~\cite{JCsurf} that this corresponds to the operator $(1,3)$ in
the Kac classification \cite{BPZ,CardyLH}, 
giving, for general $q$, $x_1=(m-1)/(m+1)$, where
$q=4\cos^2\big(\pi/(m+1)\big)$, and $x_1=\frac13$ for $q=1$. 
This conjecture agrees
with the known exact result for $q=2$ and numerical work for $q=3$ and
$q=1$. Its correctness is also born out by the numerical success of the
crossing formula of Ref.~\cite{Cardy}. We now further conjecture that
the representations $R_n$ correspond to $(1,2n+1)$ in the Kac
classification. This is based on the fusion rules for these
representations. Observe that the composition law for the $S_q$
representations under consideration 
is isomorphic to that for addition of spin $n$ in $SU(2)$,
which in turn is the same as the fusion rules \cite{BPZ} for the Kac
representations $(1,2n+1)$
in conformal field theory (in non-minimal models corresponding to
generic values of $q$). Thus, for example, insertion
of two order parameters $\varphi_a$ and $\varphi_b$ near the end of the
strip will in general give rise to propagating states corresponding to the 
tensor representation $\varphi_{ab}$ (when $a\not=b$), the vector
representation (when $a=b$) and the identity representation. Since these
last two correspond to $(1,3)$ and $(1,1)$ respectively, we may identify the
first with $(1,5)$. This argument may be generalised straightforwardly to
higher values of $n$. Then,
according to the Kac formula \cite{BPZ},the highest weight
of the $(1,2n+1)$ representation is $x_n=n(mn-1)/(m+1)$, or 
$x_n=n(2n-1)/3$ for $q=1$. This, combined with (\ref{ee}), gives the
first result (\ref{bb}), valid as $k\to\infty$ at fixed $n$. 

The large $n$ behaviour of (\ref{bb}) may also be derived directly from 
Coulomb gas arguments \cite{Nienhuis}. In this approach, the configurations
of the critical
cluster model are mapped onto those of densely packed loops
on the surrounding lattice. Each loop carries a factor $q^{1/2}$,
which may be traded for local weights by considering each loop as 
corresponding to two oriented loops with vertex weights $e^{\pm i\chi}$
according to whether they turn to the right or left at a given site,
and setting $q^{1/2}=2\cos4\chi$. These loop configurations are then
mapped onto those 
of a local height model on the dual lattice, with heights $\phi(r)\in
\frac\pi2{\bf Z}$, and the rule that 
the height difference between neighbouring dual sites is
$\pm\frac\pi2$ according to the orientation of the corresponding
dual bond. This in turn
is supposed to renormalise onto a Gaussian model with reduced hamiltonian
$(g/4\pi)\int(\nabla\phi)^2d^2\!r$, where $g=2(2-8\chi/\pi)$, and
$2\leq g\leq4$. Free boundary conditions on the original Potts model
correspond to Dirichlet conditions $\phi={\rm constant}$ in the height model.

\begin{figure}
\centerline{
\epsfxsize=5in
\epsfbox{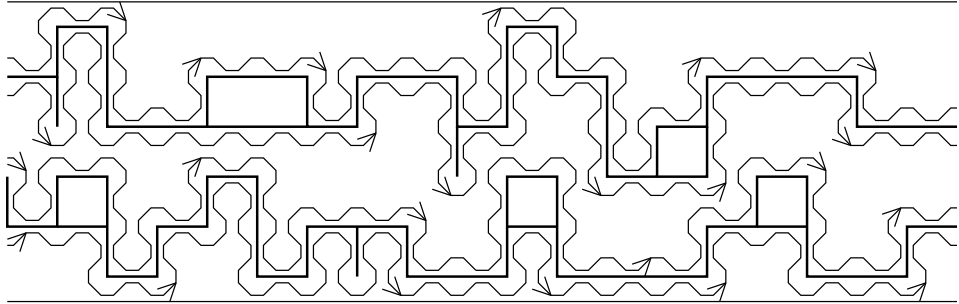}}
\caption{Configuration in which two clusters span the strip. The hulls
of these correspond to a loop configuration with charge 4. Other possible
non-spanning clusters are not shown.}
\label{2n}
\end{figure}

In the strip geometry, the total charge, that is the number of left-oriented
minus right-oriented loops, is conserved along the strip.
Consider the configurations where this charge is $2n$. These correspond to
cluster configurations where at least $n$ distinct clusters traverse the 
strip, as illustrated in Fig.~\ref{2n}. 
In the height model this means that
the difference in the heights between the upper and lower edges is fixed to
be $2n(\pi/2)=n\pi$. Neglecting fluctuations, the energy of such a
configuration is simply $(g/4\pi)(n\pi/L)^2\cdot kL\cdot L=(g\pi n^2/4)k$.
Inserting the $q=1$ value $g=\frac83$ gives the leading term in the
result (\ref{bb}) for $\ln P$. This calculation works only for large $n$
because (a) the mapping to the Gaussian model is valid only in the bulk and
not close to the boundary, and only for large $n$ are most of the loops far
from the boundary; and (b) because the fluctuations are expected to give an
$O(1)$ contribution.

A similar argument, in this case yielding the exact result at large $k$,
 may be applied to the cylindrical geometry, where there
are periodic boundary conditions around the strip. Once again, the loop 
configurations with charge $2n$ correspond to at least $n$ clusters connecting
the ends of the cylinders. Writing $\phi=n\pi v/L+\phi'$, where $v$ is the 
coordinate around the cylinder and $\phi'$ satisfies periodic boundary
conditions, the energy functional is 
$(g\pi n^2/4)k+(g/4\pi)\int(\nabla\phi')^2d^2\!r$, where the first term
is identical to that for the strip with free boundaries. The integral over
the fluctuating part then gives a contribution \cite{BCN}
$(\pi c_G/6)k$ to $\ln P$, where
$c_G=1$ is the central charge of the free scalar field $\phi'$.
Putting these together ans setting $g=\frac83$ then gives the result 
in (\ref{bbc}). Note that this is correct only at $q=1$: in general it
should be normalised by the partition function. The finite-size corrections
to this have the above form, with $c=0$ at $q=1$. This comes about because
the Gaussian result $c_G=1$ is reduced by the effects of loops which can 
wind around the cylinder \cite{Deltac}. These are forbidden when
other loops already extend along the cylinder, so that no similar reduction
occurs in this case.

In general, the behaviour in (\ref{bbc}) should be of the
form $2\pi x_n^{(b)}k$, where $x_n^{(b)}$ is 
a \em bulk \em exponent \cite{Cardy84}. In fact, these exponents are the 
so-called \em multi-hull \em scaling dimensions discussed by Duplantier
and Saleur \cite{DupSal}. In the plane, these determine the power law decay
$\sim |r_1-r_2|^{-2x_n^{(b)}}$ 
of the probability that two points $r_1$ and $r_2$ lie in the vicinity of
the external boundaries, or hulls, of $n$ distinct clusters. These are computed
in the loop gas in terms of configurations with $2n$ oriented lines running
from $r_1$ to $r_2$, which is precisely what we have argued above determines
$\ln P$ for large $k$. Our result in (\ref{bbc}) agrees with that
of Ref.~\cite{DupSal} for these exponents.

As indicated, (\ref{bbc}) does not hold for $n=1$. This is because a single
cluster which connects the ends of the cylinder is also allowed to wrap
around it: this is clearly not allowed for $n\ge2$. For $n=1$ the equivalence
to the hull exponents no longer holds. Instead, we expect for large $k$
that $P(1,k,L)$ is asymptotically equal to the probability that the two
ends are connected (by any number of clusters) and it should behave as
$\exp(-2\pi \tilde x k)$, where $\tilde x=\frac5{48}$ is the usual magnetic 
scaling dimension of the $q=1$ Potts model, which gives the probability
$\sim |r_1-r_2|^{-2\tilde x}$ that points $r_1$ and $r_2$ in the plane
are connected.  Thus for $n=1$ on the cylinder, (\ref{bbc}) is replaced by
\be\label{n1}
\ln P(1,k,L)\sim -(5\pi/24)k.
\ee
However, the result in (\ref{bbc}) with $n=1$ does have a physical meaning:
it is the asymptotic probability that the two ends of the cylinder are connected
by a cluster which does not also wrap around the cylinder. As expected,
this is much smaller then the unrestricted probability.

Finally we discuss whether it is possible to compute $P(n,k,L)$
for non-asymptotic values of $k$ and $n$, in the scaling limit $L\to\infty$. 
This corresponds to a generalisation of the calculation of
Ref.~\cite{Cardy}, and first
involves identifying suitable boundary conditions corresponding to
the states $|A\rangle$. It is not difficult to see that these states
should be suitable linear combinations of states corresponding to
boundary conditions in which each Potts spin is constrained to lie in a
subset of $n$ states out of the possible $q$. Let us denote the boundary
state in which each spin is constrained to take the values $a$ or $b$ or
$\ldots$ (where all the $a$,$b$,$\ldots$ are different)
by $|ab\ldots\rangle$. Then 
$P(n,k,L)\propto\langle A_n|e^{-kL\H(L)}|A_n\rangle$, where, for example,
\begin{eqnarray}
|A_1\rangle&=&|a\rangle-|b\rangle,\\
|A_2\rangle&=&|ab\rangle+|cd\rangle-|ac\rangle-|bd\rangle,
\end{eqnarray}
and so on.
It may be seen that that these states do indeed transform according to
the advertised representations of $S_q$, and so will couple precisely to
the representations $R_n$, and this will be the dominant coupling in the
limit $k\to\infty$.
However, in order to determine the dependence of $P(n,k,L)$
for finite $k$, in analogy with the argument of Ref.~\cite{Cardy}, it is
necessary to determine the four-point function of boundary condition
changing operators which connect the above boundary conditions to the
free boundary conditions along the other edges. Unlike the case of
Ref.~\cite{Cardy} it does not appear that these operators, for $n>1$,
correspond to simple Kac representations. However, it may still be
possible to conjecture a suitable differential equation or an integral
representation for this function, as was done recently by
Watts \cite{Watts} in the case of the probability of a simultaneous
left-right and up-down crossing of the rectangle.
A simpler case to consider might be that of clusters which span a 
cylinder of finite length. It is known that in this case the 
appropriate matrix elements may be expressed as a linear combination 
of Virasoro characters \cite{CardyB}. 

We note that the simple argument given above in (\ref{pp})
(and the rigorous arguments of Aizenman \cite{A2}) provide a simple
physical reason why the scaling dimensions of composite operators such as
those discussed should increase like $n^2$ in two dimensions. 
The generalisation of Aizenman's argument to $d$ dimensions suggests
that the rate of increase of $-\ln P(n,k,L)$ is like $n^{d/(d-1)}$. 
However, for
$d>2$ this quantity is no longer related to scaling dimensions by
conformal invariance.

Our conjecture that the relevant scaling dimensions in (\ref{bb}) 
are the $(1,2n+1)$ operators in the Kac classification is equivalent to
a result of Saleur and Bauer \cite{SB} for the spin-$n$ operators in
the Bethe ansatz solution of the equivalent vertex model. These are the
`boundary multi-hull' operators.

After this work was completed, we saw the paper of 
Shchur and Kosyakov \cite{SK}, which reports Monte Carlo measurements
of $P(n,1,L)$ for $n=2$ and $n=3$ on lattices with $L$ up to 64.
Their quoted results agree well with our predictions in (\ref{bb},
\ref{bbc}), even though the value of $k=1$ is not large. In particular
the ratios of $-\ln P(n,1,L)$ between the cases of open boundaries (\ref{bb})
and periodic boundary conditions (\ref{bbc}) is predicted to be 
$\frac45=0.8$ for $n=2$ and $\frac67\approx0.857$ for $n=3$. The 
corresponding values quoted in Ref.~\cite{SK} are $0.808(10)$
and $0.851(20)$. This close agreement with the asymptotic form may
be explained by the observation that the higher eigenstates of $\H$ in
(\ref{dd}) give corrections of order $e^{-2\pi k}$.
For $n=1$, using (\ref{n1}) we find a ratio $\frac85=1.6$, to be
compared with the value $1.5348$ extracted from the result $P(1,1,L)\approx
0.63665(8)$ of Hovi and Aharony \cite{HA} for the cylinder, and the
exact result of $0.5$ for the square.

The author thanks M.~Aizenman, H.~Saleur and L.~N.~Shchur
for useful correspondence
and discussions. This work was completed while the author was visiting
the Institute for Theoretical Physics, Santa Barbara, and was 
supported in part by the Engineering and
Physical Sciences Research Council under Grant GR/J78327, and by the
National Science Foundation under Grant PHY94-07194.

\end{document}